\documentclass[11pt]{article}\usepackage{amsmath,amssymb,multirow,jheppub,graphicx}
\usepackage{centernot}
\usepackage{color}
\allowdisplaybreaks[1]
\usepackage{tikz}
 \tikzset{node distance=2cm, auto}
 \usepackage{textcomp}
 \usepackage{array}
 \usepackage{xspace}
\usepackage{fancyvrb}

\usepackage{verbatim}   
\usepackage[all]{xy}
\usepackage{bm}  
\def\Dslash{{\rlap{\raise 1pt \hbox{$\>/$}}D}}
\def\Pslash{{\rlap{\raise  1pt \hbox{$\>/$}}\,\partial}}

\newcommand{\be}{\begin{equation}}      
\newcommand{\ee}{\end{equation}}      
\newcommand{\bea}{\begin{eqnarray}}      
\newcommand{\eea}{\end{eqnarray}}

\newcommand{\Mathematica}{{{\scshape Mathematica}\textsuperscript{\textregistered}}}
\newcommand{\BenderWu}{\texttt{BenderWu}}
\newcommand{\BWProcess}{\texttt{BWProcess}}

\newcommand{\ket}[1]{\left|#1\right\rangle}
\newcommand{\bra}[1]{\left\langle#1\right|}

\usepackage[T1]{fontenc}
\usepackage{ucs}
\usepackage{times}

\title{
Aspects of Perturbation theory in Quantum Mechanics:  The \texttt{BenderWu} \Mathematica{} package}
\author{Tin Sulejmanpasic} \author{and Mithat  \"Unsal}
\affiliation{Department of Physics, North Carolina State University, Raleigh, NC 27695, USA}
\emailAdd{tin.sulejmanpasic@gmail.com}
\emailAdd{unsal.mithat@gmail.com}
\abstract{ 
    }

\abstract{We discuss a general setup which allows the study of the perturbation theory of an arbitrary, locally harmonic 1D quantum mechanical potential as well as its multi-variable (many-body)  generalization.  The latter may form a prototype for regularized quantum field theory.
We first  generalize the method of Bender-Wu, and derive  exact recursion relations which allow the determination of the perturbative wave-function and energy corrections to an arbitrary order, at least in principle. 
   For 1D systems,  we implement these equations in an easy to use \Mathematica{} package we call \BenderWu{}. Our package enables quick home-computer computation of high orders of perturbation theory  (about  100 orders in   10-30 seconds, and  250 orders in 1-2h)  and enables practical study of a large class of problems in Quantum Mechanics. 
   We have two hopes concerning the     \BenderWu{} package. One is that due to resurgence, large amount of  non-perturbative information, such as non-perturbative energies and wave-functions (e.g. WKB wave functions), can in principle be extracted from the perturbative data. We also hope that the package may be used as a teaching tool, providing an effective bridge  between  perturbation theory and non-perturbative physics in textbooks. 
Finally, we show that for the multi-variable case, the recursion relation acquires a geometric character, and has a structure which allows parallelization to computer clusters. 

\vspace{.5cm}
{\small\scshape\noindent Bundled with the source files of this document is a folder ``\texttt{BenderWu\_v1.0}'' which contains the \Mathematica{} package files and instructions. The most up-to-date package files can be accessed at \url{http://library.wolfram.com/infocenter/MathSource/9479/}}
}

\begin{document}
\maketitle
\newpage

\section{Introduction}

Time-independent perturbation theory in quantum mechanics, developed by Erwin Schr\"odinger, is almost as old as quantum mechanics itself. Published in 1926 \cite{schrodinger_pert_th}, the same year as the Schr\"odinger equation \cite{PhysRev.28.1049}, it is a standard topic of any  textbook of quantum mechanics. However the method is rooted in wave-mechanics and dates back to Lord Rayleigh and his 1877 textbook on \emph{The Theory of Sound} \cite{rayleigh1877}. For this reason the theory is often referred to as \emph{ The Rayleigh-Schr\"odinger perturbation theory.}

Little did these great minds know that had they been able to compute large orders of perturbation theory, for most systems they would have come to a surprising  revelation: the radius of convergence of the perturbation theory is zero. The reason for this is a factorial growth of the coefficient of the expansion parameter, which we refer to here as ``the coupling'' and denote  by $g$. 

Rather than being a curse, it is now becoming increasingly clear that the perturbative expansion is intimately tied up with non-perturbative physics. The analysis and study of such series, functions and theories goes under the name of \emph{Resurgence theory}. Although the resurgence idea is old \cite{Ecalle-book}, it has been the subject of recent rapid development, from one-dimensional integrals \cite{Berry657,Howls,delabaere2002}, through quantum mechanics \cite{Bogomolny:1980ur, ZinnJustin:1981dx, ZinnJustin:1982td,Dunne:2013ada,Dunne:2014bca,Misumi:2015dua} to quantum field theory \cite{Dunne:2012ae,Cherman:2014ofa}  and string theory, see e.g. \cite{Marino:2008ya,Garoufalidis:2010ya,Aniceto:2011nu,Aniceto:2013fka}.  

The resurgence structures are mostly studied in Quantum Mechanics. However, to our knowledge, all publications to date deal with case-by-case examples, and no practical general procedure of studying large orders of perturbation theory was published. In this work we adapt the method, developed originally by C.~M. Bender and T.~T. Wu \cite{Bender:1990pd} for the anharmonic oscillator with quartic term in the potential, to a perturbative expansion of an \emph{arbitrary locally-harmonic potential} around one of its harmonic minima\footnote{The procedure discussed here is most likely applicable to any system where the leading order of the potential can be computed analytically. An interesting practical system to adapt to this method would be to the case of a coulomb potential, which is of importance in atomic physics and chemistry.}.
In addition we develop a workable \Mathematica{}  computer code which can easily compute many orders of perturbation theory (about $\sim100$ orders in $\sim 10-30$ seconds, and $\sim250$ orders in  $1-2$h on a modern home computer). The computation is done symbolically by default, so the result is an exact result in perturbation theory without any numerical error. Furthermore, the potential can be allowed to depend on arbitrary symbolic  variables, allowing the study of the parametric dependance of the perturbation theory.

To be more specific,  the code presented here allows a perturbative treatment of a one-dimensional systems with a Hamiltonian in the coordinate representation given by
\be
H=-\frac{\hbar^2}{2m}\frac{\partial^2}{\partial X^2}+\mathcal V(X)\;,
\ee
where $\mathcal V(X)$ is an arbitrary non-singular potential. Such a potential is typically characterized by some length scale $a$ characterizing its spatial variation, and a frequency scale of the harmonic motion around one of its minima\footnote{The frequency may be different at individual minima, but the rate of the particle oscillations around them is set only by a single frequency scale $\omega$. Note that in subsequent sections we will define $\omega$ to be the dimensionless number specifying the coefficient of the quadratic term of the $v(X/a)$.}. On dimensional grounds we can write in general
\be
\mathcal V(x)=m\omega^2 a^2v(X/a)\;,
\ee
where $v(X/a)$ is an arbitrary (dimensionless) function defining a non-trivial potential. The  perturbation theory is defined by an expansion of the potential $\mathcal V(X)={m\omega^2}{a^2}v(X/a)$ around one of its  local minima, which coincide with the minima of $v(X/a)$. Without loss of generality, we will take that one such minimum is at $x=0$, and construct the perturbation theory around it. A dimensionless combination of parameters is given by\footnote{The square root is inserted for convenience, because, as we shall see, the wave-function is in general an expansion in $g$ defined in this way. The energy eigenvalue, however, will be an expansion in $g^2$.} $g=\sqrt{\frac{\hbar}{m\omega a^2}}$, which is made small when either $\omega, a$ or $m$ is made large. A time-independent perturbation series is therefore an expansion of the wave-function and energy in this small coupling, by an expansion of the potential $v(X/a)$ in powers of $X/a$, treating non-quadratic terms as a perturbation of a Harmonic oscillator.

To make this more explicit, let us go to a dimensionless spatial coordinate\footnote{The factor of $g$ is inserted for convenience.} $x=X/(ag)$.  Then the  Hamiltonian becomes
\be
H=\frac{\hbar^2}{m a^2g^2}\left[-\frac{1}{2}\frac{\partial^2 }{\partial  x^2}+\frac{v( x{g})}{g^2}\right]\;.
\ee
Finally, we define a reduced Hamiltonian given by
\be\label{eq:red_hamiltonian}
h=\frac{ma^2g^2}{\hbar^2}H=-\frac{1}{2}\frac{\partial^2 }{\partial x^2}+\frac{v(xg)}{g^2}\;.
\ee
We wish to construct the perturbative solution to the Schr\"odinger equation by expanding the potential $v(xg)$ for small $g$ around one of its minima, taken to be at $x=0$,
\be
\frac{v(xg)}{g^2}=\frac{v(0)}{g^2}+\frac{1}{2}v''(0)x^2+\frac{g}{3!}v'''(0)x^3+\dots\;.
\ee
It is clear that the higher terms are made arbitrarily small by taking $g$ arbitrarily small. We can therefore treat all terms, except the quadratic term, as a perturbation, and solve the reduced Schr\"odinger equation
\be\label{eq:red_schro}
h\psi(x)=\epsilon \psi(x)
\ee
for $\psi(x)$ and $\epsilon$ as a power series 
\be
\psi(x)=\psi_0(x)+\psi_1(x){g}+\psi_2(x)g^2+\dots
\ee
and\footnote{Notice that if $v(0)\ne 0$, then there will be a $1/g^2$ energy contribution. We will almost entirely ignore this classical shift, and even suppress its output by default in the \BenderWu{} package. }
\be
\epsilon=\epsilon_0+\epsilon_1 g+\epsilon_2 g^2+\dots\;.
\ee
Note however that $\epsilon_{2n+1}=0$ (see \ref{sec:symmetry}) for the potential at hand.
To do this we will apply the Bender-Wu method, originally developed for an anharmonic oscillator with the $gx^4$ perturbation \cite{Bender:1990pd}. As we shall see, however, it is possible to generalize this method to an arbitrary potential, as long as its minima are harmonic and non-singular. 

We will also be interested in a more general problem with the reduced Hamiltonian
\be\label{eq:gen_schrodinger}
h=-\frac{1}{2}\frac{\partial^2}{\partial x^2}+V(x)
\ee
where $V(x)$ is given by
\be\label{eq:ferm_pot}
V(x)=\frac{v_0(xg)}{g^2}+v_2(xg)\;.
\ee 
The term $v_2(x)$ is clearly suppressed by a power of $g^2$ compared to the first term, and is important in the definition of supersymmetric quantum mechanics, as well as some quasi exactly solvable models which were an initial motivation for this work. Studies of these problems is the topic of a separate publication \cite{CanYuya}. Note however that such potentials cannot be classical in nature, and must come from some quantum effects. 

Although we will not implement it in the \Mathematica{} package which accompanies this work, we also give \BenderWu{} recursion relations for the potential of the form
\be\label{eq:pot_gen_form}
V(x)=\frac{1}{g^2}\sum_{m=0}^{\mathcal N}{g^m}{v_m(xg)}
\ee
for an arbitrary integer $\mathcal N$. This potential is of some interest in the literature (see for example a elegant book \cite{kawai2005algebraic}).  

Finally we also note the multi-variable (many-body)  Schr\"odinger problem
\be
\left(-\sum_{i=1}^{N}\frac{1}{2}\partial_i^2+\frac{1}{g^2}v(gx_1,gx_2,\dots,gx_N)\right)\psi=\epsilon\psi\;.
\label{multi-var}
\ee
is possible to treat by our methods in principle. This problem, although a natural generalization of the one-variable problem, is not pertinent to the rest of the paper. Nevertheless, given the importance of such a problem for higher dimensional and multi-particles QM problems, as well as its extensions to quantum field theory\footnote{Quantum field theory can be thought of as  quantum mechanics  where the number of degrees of freedom is sent to infinity.}, we felt compelled to derive the recursion relations for \eqref{multi-var} and explain how they are solved, at least in principle. We also restrict the discussion to non-degenerate perturbation theory only, leaving the study of general degenerate case for the future. The efficacy of this approach in the multi-dimensional problem above is also left for future work.

The paper is organized as follows: In Section \ref{sec:BW} we generalize the method of Bender-Wu to the arbitrary potential in 1D quantum mechanics. In Section \ref{sec:multi}, we sketch how the discussion generalizes to the multi-variable quantum mechanics, and briefly discuss the geometric structure which arises there. Section \ref{sec:BenderWu} is dedicated to the explanation of the \BenderWu{}  \Mathematica{} package. The content of this section is largely independent of the rest of the paper, and the user interested in learning how to run and use the package is invited to go there immediately. We conclude in Section \ref{sec:conclusions}.


\subsection*{Comments on the Literature}

We dedicate this small section to briefly comment on the literature available and the application of the Bender-Wu method. Following the original publication the vast application of the method was used to analyze the anharmonic oscillator. A notable exception is \cite{Stone:1977au}, which attempted to generalize the method to a Mathieu potential, but encountered a numerical instability of the recurrence relations for level numbers $\nu>1$. Reference \cite{Verbaarschot:1990ga} applied the method to the supersymmetric version of the double-well potential. In \cite{Bender:1998ry} the method was applied to a PT symmetric potential $x^2+igx^3$, while in \cite{Bender:2001ap} the complex H\'enon-Heiles potential was analyzed. All of these results, save the last one\footnote{The potentials considered in \cite{Bender:2001ap} involve two and three dimensional potentials, which are currently not implemented in the \BenderWu{} package. See however Section \ref{sec:multi} however.}, are easily reproduced the \BenderWu{} function of our package, within seconds.


\section{The Bender-Wu method for arbitrary locally harmonic potentials in one dimension}\label{sec:BW}

In this section we will develop the recursion relations which will allow us to compute the perturbative expansion of the wave-function and energy. The method goes under the name of Bender-Wu who first invented it to study the quartic potential problem in quantum mechanics \cite{Bender:1990pd}. Here we adapt the method for a general classical potential with the structure $v(gx)/g^2$,  and some of its generalization to quantum effective potentials. It is important to realize that the potential can even be a periodic function, or, indeed, any function with a harmonic minimum at $x=0$, and infinitely differentiable at $x=0$. In the next section we will also consider adding the potential of the form \eqref{eq:pot_gen_form}.

The contents of this section are not important for understanding how to use the \BenderWu{} package and its functions. A reader who is not interested in details of the generalized Bender-Wu method is encouraged to proceed to Sec.~\ref{sec:BenderWu} as well as to the installation and example \Mathematica{} notebooks accompanying this work.

\subsection{The classical 1D potential}\label{sec:BWclassical}

Consider the Schr\"odinger equation
\be\label{eq:gen_schro}
-\frac{1}{2}\psi''(x)+\frac{1}{g^2}v(gx)\psi(x)= \epsilon\psi(x)\;.
\ee
We wish to construct a perturbative expansion around a harmonic minimum which, without loss of generality we take to be located at $x=0$ and set\footnote{This corresponds to a classical shift in energy and can simply be reinstated by introducing a contribution $\epsilon_{-2}=v(0)$.}  $v(0)=0$. We first set
\be
\psi(x)=u(x) e^{-\frac{\omega x^2}{2}}\;,
\ee
where\footnote{Note the difference between the definition of $\omega$ here and the introduction.}
\be
\omega^2=v''(0)\;.
\ee
which reduces the Schr\"odinger equation to
\be\label{eq:ueq}
-u''(x)+2\omega x u'(x)+\frac{2}{g^2}\tilde v(gx)u(x)=2\left(\epsilon-\frac{\omega}{2}\right)u\;.
\ee
Here we defined
\be
\tilde v(g x)/g^2=v(gx)/g^2-\frac{\omega^2}{2}x^2\;.
\ee
Now let us write a formal expansion of $u(x)$ in powers of $g$.
\be
u(x)=\sum_{l=0}^\infty u_l(x) g^l\;.
\ee
Further we expand
\be
\tilde v(gx)/g^2=\sum_{n=1}^\infty v_n g^n x^{n+2}\;, \qquad v_n=\frac{1}{(n+2)!}V^{(n+2)}(0)\;.
\ee
Now we formally write an expansion for $\epsilon$
\be
 \epsilon=\sum_{n=0}^\infty \epsilon_n g^n\;.
\ee
Plugging these into the eq. \eqref{eq:ueq} and equating the powers of $g$, we have
\be\label{eq:ul-eq}
-u_{l}''(x)+2\omega x u_l'(x)+2\sum_{n=1}^l v_{n}x^{n+2}u_{l-n}(x)=2\sum_{n=0}^l(\epsilon_n -\delta_{n0}\omega/2)u_{l-n}(x)
\ee
Notice that the leading order contribution is given by $l=0$, for which
\be
-u_0''(x)+2\omega xu_0'(x)=2(\epsilon_0-\omega/2) u_{0}(x)\;,
\ee
which is the Hermite equation, so that $u_0(x)=H_\nu(x)$ and $\epsilon_0-\omega/2=\nu \omega$, where $\nu$ is the \emph{level number}. This in turn gives the leading order energy\footnote{Keep in mind that in principle $\psi,\epsilon,$ and therefore $u_l(x)$ and $\epsilon_l$ also depend on $\nu$, which defines the level number. To avoid unnecessary clutter, we make this dependence implicit, setting $\nu$ to a fixed integer. 
}
\be
\epsilon_0=\omega\left(\nu+\frac{1}{2}\right)\;,
\ee 
as expected. 

Now let us expand $u_l(x)$ in powers of $x$
\be
u_l(x)=\sum_{k=0}^{K_l}A_l^k x^k\;,
\ee
where $K_l$ is the maximal power of $x$ that can appear at order $l$. The upper bound of $K_l$ is given by an expression\footnote{The exact expression for $K_l$ is given by \eqref{eq:K_l_formula} in Appendix \ref{app:highest_power}, which clearly obeys this inequality. The upper bound however will suffice for solving the recursion relations. }
\be\label{eq:K_l}
K_l\le \nu+3l\;,
\ee
where the equality holds if $v_1\ne0$. Plugging into \eqref{eq:ul-eq} and equating powers of $x$, we have
\be\label{eq:Arec}
-(k+2)(k+1)A_l^{k+2}+2\omega k A_l^{k}-2\omega \nu A_l^k=2\sum_{n=1}^l \left(\epsilon_n A_{l-n}^k-v_n A_{l-n}^{k-n-2}\right)
\ee

The equations \eqref{eq:Arec} can now be solved recursively for $A_{l}^k$ and $\epsilon_l$. We proceed in the following way. First notice that if $k=\nu$, rearranging \eqref{eq:Arec} we have
\be\label{eq:eps-l_temp}
2\epsilon_{l}A_0^{\nu}=-(\nu+2)(\nu+1)A_l^{\nu+2}-2\sum_{n=1}^{l-1}\epsilon_nA_{l-n}^\nu+2\sum_{n=1}^l v_n A_{l-n}^{\nu-n-2}
\ee
So we can determine the energy $\epsilon_l$ as long as we have all $A$-coefficients with order smaller than $l$ and as long as we find the coefficient $A_l^{\nu+2}$.
Further we choose 
\be\label{eq:A_norm}
A_0^\nu=1\;, A_l^\nu=0, \forall l>0
\ee
as a normalization\footnote{This just fixes the coefficient of $x^\nu$ in the wave-function to be unity. Other normalizations will give the same result for all physical observables.}, so that \eqref{eq:eps-l_temp} becomes 
\be\label{eq:eps-l}
\epsilon_{l}=-\frac{1}{2}(\nu+2)(\nu+1)A_l^{\nu+2}+\sum_{n=1}^l v_n A_{l-n}^{\nu-n-2}
\ee
Next we rearrange the equation \eqref{eq:Arec} to read
 \be\label{eq:Arec-1}
 A_{l}^k=\frac{1}{2\omega(k-\nu)}\left[(k+2)(k+1)A_l^{k+2}+2\sum_{n=1}^{l}\epsilon_nA_{l-n}^k-2\sum_{n=1}^{l}v_n A_{l-n}^{k-n-2}\right]\;.
 \ee

We are now in a position to use \eqref{eq:Arec-1} and \eqref{eq:eps-l} to solve for all $A_{l}^k$ and $\epsilon_l$ recursively. The steps are as follows:
\begin{itemize}
\item Consider  the equation \eqref{eq:Arec-1} first for\footnote{Recall that for $l=0$ $A_{l}^k=0$ for $k>\nu$ as the polynomial $u_0(x)$ are just Hermite polynomials.} $k>\nu$ and $l>0$, giving

 \[
 A_{l}^k=\frac{1}{2\omega(k-\nu)}\left[(k+2)(k+1)A_l^{k+2}+2\sum_{n=1}^{l-1}\epsilon_nA_{l-n}^k-2\sum_{n=1}^{l-1}v_n A_{l-n}^{k-n-2}\right]\;.
 \]
We emphasize that the $\epsilon_l$ does not appear on the right hand side because in \eqref{eq:Arec-1} it multiplies a coefficient $A_{0}^k$, which is zero for $k>\nu$. Furthermore, since $A_l^{K_l+2}=0$, by taking $k=K_l$ we can compute $A_l^{K_l}$, as well as $A_l^{K_l-1}$ if we know all coefficients and energies of order $<l$. Now that we know $A_l^{K_l}$ and $A_l^{K_{l}-1}$, we can compute $A_l^{k}, k=K_l,K_l-1,K_l-2,\dots \nu+1$. Notice that an overestimate  in $K_l$ does not invalidate this procedure, so, by  \eqref{eq:K_l}, we may as well start with $k=\nu+3l$ and solve in sequence for $k=\nu+3l,\nu+3l-1,\dots, \nu+1$. 

\item Now that all $A_l^k,k>\nu$ are known, and in particular $A_l^{\nu+2}$ is known, use \eqref{eq:eps-l}
\[
\epsilon_{l}=-\frac{1}{2}(\nu+2)(\nu+1)A_l^{\nu+2}+\sum_{n=1}^l v_n A_{l-n}^{\nu-n-2}\;.
\]
to compute $\epsilon_l$.
\item Finally, now that we know $\epsilon_l$ we use \eqref{eq:Arec-1} again to obtain the remaining $A_l^{k}$ with $k<\nu$ (recall that $A_l^{\nu}=0, l>0$ by our normalization). 
 \[
 A_{l}^k=\frac{1}{2\omega(k-\nu)}\left[(k+2)(k+1)A_l^{k+2}+2\sum_{n=1}^{l}\epsilon_nA_{l-n}^k-2\sum_{n=1}^{l-1}v_n A_{l-n}^{k-n-2}\right]\;.
 \]
\end{itemize}
Next we will discuss the generalized problem \eqref{eq:gen_schrodinger} with $V(x)$ given by \eqref{eq:ferm_pot}.

\subsection{Adding the quantum effective action}\label{sec:QEA}

The discussion so far only considered classical potentials, not containing powers of the Planck constant. However, as was mentioned in the introduction, we are also interested in describing supersymmetric problems, as well as so-called Quasi-Exactly Solvable (QES) problems which are of the form
\be
\left(-\frac{1}{2}\partial_x^2+V(x)\right)\psi=\epsilon\psi\;,
\ee
with
\be\label{eq:bos_plus_ferm_pot}
V(x)=\frac{1}{g^2}v_0(gx)+v_2(gx)\;.
\ee
Further if we define $v_n^0$ and $v_n^2$ as coefficients of expansion
\begin{align}
&\omega=\sqrt{v_0''(0)}\;,\\
&v_0(gx)/g^2-\omega^2x^2/2=\sum_{n=1}^\infty g^n x^{n+2} v_n^0\;, &&v_n^b=\frac{v_0^{(n+2)}(0)}{(n+2)!}\\
&v_2(gx)=\sum_{n=0}g^n x^n v_n^2\;,  &&v_n^f=\frac{v_2^{(n)}(0)}{n!}
\end{align}
Most of the previous discussion remains unchanged, but relation \eqref{eq:Arec} now becomes
\be\label{eq:Arec-ferm}
-(k+2)(k+1)A_l^{k+2}+2\omega k A_l^{k}-2\omega \nu A_l^k=\sum_{n=1}^l \left(2\epsilon_n A_{l-n}^k-2v_n^b A_{l-n}^{k-n-2}-2v_n^f A_{l-n}^{k-n}\right)
\ee
while \eqref{eq:eps-l} and \eqref{eq:Arec-1} become
\begin{align}\label{eq:eps-l_ferm}
&2\epsilon_{l}A_0^{\nu}=-(\nu+2)(\nu+1)A_l^{\nu+2}-2\sum_{n=1}^{l-1}\epsilon_nA_{l-n}^\nu+2\sum_{n=1}^l (v_n^b A_{l-n}^{\nu-n-2}+v_n^f A_{l-n}^{\nu-n})\\
\label{eq:Arec-1_ferm}
 &A_{l}^k=\frac{1}{2\omega(k-\nu)}\left[(k+2)(k+1)A_l^{k+2}+\sum_{n=1}^{l}2\epsilon_nA_{l-n}^k-2\sum_{n=1}^{l}(v_n^b A_{l-n}^{k-n-2}+v_n^f A_{l-n}^{k-n})\right]\;.
 \end{align} 

 In particular we will choose the normalization which sets the coefficient of $x^{\nu}$ to unity, so that $A_0^\nu=1$ and $A_0^k=0,k\ne\nu$. This turns \eqref{eq:eps-l_ferm} into
 \be\label{eq:eps-l_ferm_norm}
 \epsilon_{l}=-\frac{1}{2}(\nu+2)(\nu+1)A_l^{\nu+2}+\sum_{n=1}^l (v_n^b A_{l-n}^{\nu-n-2}+v_n^f A_{l-n}^{\nu-n})
 \ee
 
\noindent The procedure for solving the recursion relations at order $l$ is identical to that described below \eqref{eq:Arec-1}, however, considering its centrality to the procedure and adhering to an adage \emph{repetitio est mater studiorum}, it is repeated here for convenience and clarity:
 \begin{itemize}
 \item Starting from $k=\nu+(3+L)l$ solve
 \[
 A_{l}^k=\frac{1}{2\omega(k-\nu)}\left[(k+2)(k+1)A_l^{k+2}+\sum_{n=1}^{l-1}2\epsilon_nA_{l-n}^k-2\sum_{n=1}^{l}(v_n^b A_{l-n}^{k-n-2}+v_n^f A_{l-n}^{k-n})\right]
 \]
 for all $k=\nu+(3+L),\nu+(3+L)-1,\dots,\nu+1$ in terms of wave-function coefficients $A_{l'}^k$ and $\epsilon_{l'}$ with $l'<l$ which are already known.
 \item Solve for $\epsilon_l$ using (recall that $A_l^\nu=0$ for $l\ne 0$ by normalization)
\[
 \epsilon_{l}=-\frac{1}{2}(\nu+2)(\nu+1)A_l^{\nu+2}+\sum_{n=1}^l (v_n^b A_{l-n}^{\nu-n-2}+v_n^f A_{l-n}^{\nu-n})
 \]
 \item Finally solve for $A_l^k$ with $k=\nu-1,\nu-2,\dots,0$ (notice the dependance on $\epsilon_l$ which was found in the previous step)
  \[
 A_{l}^k=\frac{1}{2\omega(k-\nu)}\left[(k+2)(k+1)A_l^{k+2}+\sum_{n=1}^{l}2\epsilon_nA_{l-n}^k-2\sum_{n=1}^{l}(v_n^b A_{l-n}^{k-n-2}+v_n^f A_{l-n}^{k-n})\right]
 \]
 \end{itemize}
 It is this algorithm that is the essence of the \BenderWu{} function contained in the \BenderWu{} \Mathematica{} package. Next we discuss the recursion relations for the general case of \eqref{eq:gen_schrodinger} with the potential \eqref{eq:pot_gen_form}. Although the \Mathematica{} package does not implement this generalization at the moment, since it is of some importance in the literature \cite{kawai2005algebraic}, and since its implementation is a trivial generalization of what has been discussed so far, we dedicate a brief statement of the recursion relations.
 
 \subsubsection{The generalization of the recursion relations to an arbitrary effective potential}
 Although we will not implement it in the \Mathematica{} code which accompanies this work, we also generalize the Bender-Wu to the potential of the form
\be
\left(-\frac{1}{2}\partial_x^2+V(x)\right)\psi=\epsilon\psi\;,
\ee
with
 \be
V(x)=\frac{1}{g^2} \sum_{m=0}^{\mathcal N} {g^m}{v_{m}(xg)}\;,
 \ee
 and define
\begin{align}
&\omega=\sqrt{v_0''(0)}\;,\\
&v_0(gx)/g^2-\omega^2x^2/2=\sum_{n=1}^\infty g^n x^{n+2} v_n^{(0)}\;, &&v_n^{(0)}=\frac{v_0^{(n+2)}(0)}{(n+2)!}\\
&v_m(gx)=\sum_{n=0}g^n x^n v_n^{(m)}\;,  &&v_n^{(m)}=\frac{v_m^{(n)}(0)}{n!}
\end{align}
while the recursion relations, appearing in the itemized list under equations \eqref{eq:Arec-1} and \eqref{eq:eps-l_ferm_norm} become
\begin{align}\label{eq:eps-l_gen}
&2\epsilon_{l}A_0^{\nu}=-(\nu+2)(\nu+1)A_l^{\nu+2}-2\sum_{n=1}^{l-1}\epsilon_nA_{l-n}^\nu+2\sum_{n=1}^l \sum_{m=0}^{\mathcal N}v_n^{(m)} A_{l-n}^{\nu-n-2+m}\\
\label{eq:Arec-1_gen}
 &A_{l}^k=\frac{1}{2\omega(k-\nu)}\left[(k+2)(k+1)A_l^{k+2}+\sum_{n=1}^{l}2\epsilon_nA_{l-n}^k-2\sum_{n=1}^{l}\sum_{m=0}^{\mathcal N}v_n^{(m)} A_{l-n}^{k-n-2+m}\right]\;.
 \end{align}

 \subsection{The symmetries and the form of the perturbation series}\label{sec:symmetry}
 
 We have already mentioned in the introduction that the corrections odd in $g$ of the energy series vanishes. This in fact holds if the potential is symmetric under a simultaneous interchange of $x\rightarrow -x,g\rightarrow -g$, which we will call $g$-parity.

Consider the reduced Schr\"odinger equation \eqref{eq:red_schro}
\be
h(x,g)\psi(x,g)=\epsilon(g)\psi(x,g)\;,
\ee
where we made explicit the dependance of $x$ and $g$ of $h,\epsilon$ and $\psi$. By a $g$-parity transformation $x\rightarrow -x,g\rightarrow -g$, the above equation maps to
\be
h(x,g)\psi(-x,-g)=\epsilon(-g)\psi(-x,-g)\;.
\ee
Therefore, $\psi(-x,-g)$ is also an eigenfunction of $x,g$ with an eigenvalue $\epsilon(-g)$. However, consider the overlap
\be
\int dx\; \psi(-x,-g)^*\psi(x,g)\;.
\ee
If the above overlap is zero, then the two states $\psi(x,g),\psi(-x,-g)$ are distinct states. This however cannot be, as making an expansion in small coupling $\psi(-x,-g)=\psi_0(-x)+\psi_1(-x)g+\dots$, where $\psi_0(-x)$ is proportional to the harmonic oscillator wave-function, and is therefore even. To leading order, the overlap is therefore nonzero. Since the two states $\psi(x,g)$ and $\psi(-x,-g)$ are the same to leading order in g, they must be so to sub-leading orders as well, therefore we have
\begin{align}
&E(g)=E(-g)\;\\
&\psi(x,g)=C(g)\psi(-x,-g)\;.
\end{align}
Notice that while the first equation implies that $E(g)$ is an even function of $g$, the wave-function in general gets multiplied by a constant which is $g$ dependent. However, we can always choose $C(g)$ to be $\pm 1$. The choice is arbitrary and it is purely a normalization issue.

In any event, while this consideration implies that the power series of energy is always in even  powers of $g$, the wave-function can always be normalized so that it contains only  powers $g^lx^k$, such that $l+k$ is even (i.e. $C(g)=1$) or $l+k$ is odd (i.e. $C(g)=-1$). 

We have derived \eqref{eq:Arec-1_ferm} and \eqref{eq:eps-l_ferm_norm} with the normalization $A_0^\nu=1$ and $A_l^{\nu}=0$, which sets $C(g)=1$ if $\nu$ is even and $C(g)=-1$ if $\nu$ is odd. The consequence of $g$-parity is that $A_l^k=0$ if ${l+k+\nu}$ is odd and $\epsilon_{l}=0$ if $l$ is odd.  Indeed we can see that if we assume that $A_{l'}^k=0$ when ${k+l'+\nu}$, and $\epsilon_{l'}=0$ when $l'$ is odd for all $l<l'$, we can see that \eqref{eq:Arec-1_ferm} implies that all $A_l^k$ will be zero if $l+k+\nu$ is odd. To see that note that if we set $l+k+\nu$ to an odd number, sums on the RHS of \eqref{eq:Arec-1_ferm} involve only coefficients $A_{l-2p}^k$, $A_{l-n}^{k-n-2}$ and $A_{l-n}^{k-n}$, for integer $p$ and $n$. However, if $l+k+\nu$ is odd, it is easily seen that all of these coefficients vanish\footnote{This is because when $l+k+\nu$ is odd, then $l-2p+k+\nu$, $l-n+k-n-2+\nu$ and $l-n+k-n+\nu$ are all odd, and so  $A_{l-2p}^k$, $A_{l-n}^{k-n-2}$ and $A_{l-n}^{k-n}$ vanish by assumption.}. Hence $A_l^{k}$ must vanish if $l+k+\nu$ is odd. Taking $l$-odd in \eqref{eq:eps-l_ferm_norm}, we see that the RHS contains terms $A_l^{\nu+2}$, $A_{l-n}^{\nu-\nu-2}$ and $A_{l-n}^{\nu-\nu}$, all of which vanish for $l$-odd. Hence $\epsilon_l=0$ with $l$--odd
 
\section{Multi-variable case}\label{sec:multi}

Although this work is mostly about the one-dimensional quantum mechanics, we wish  to present  a generalization to multi-variable case. The content of this section is not implemented in the \BenderWu{}  package,  and can be skipped entirely without affecting the understanding of the remainder of the paper. Nevertheless the study of this case is important for multiple reasons, some of which we emphasize here
\begin{itemize}
\item The study of of higher dimensional quantum mechanics systems,
\item The study of many-body systems,
\item The study of quantum-field theories as a limit when number of degrees of freedom goes to infinity.
\end{itemize}

We wish to emphasize the final point that the discussion contained in this section may have direct application to quantum field theory, and the role of resurgence there. A skeptical reader may have doubts that the limiting process of infinitely many degrees of freedom may invalidate all of the resurgence structure that is believed to hold in quantum mechanics. Indeed for a long time it was commonly believed that the IR renormalon singularities, first discussed by t' Hooft \cite{'tHooft:1977am}, cannot be fixed by any resurgence-like  perturbative/non-perturbative interplay\footnote{A progress in this direction is provided in 2d and 4d QFTs  \cite{Argyres:2012ka,Argyres:2012vv,Dunne:2012ae}, also see \cite{Cherman:2013yfa,Anber:2014sda}. }.  Our perspective is that 
the  difference between quantum mechanics and quantum field theory is not a major obstacle. Regularized QFT with 
the  UV and IR cutoffs can be interpreted as  a many-body quantum mechanics. In this respect, our proposal  is similar to lattice field theory, and may provide an alternative thereof.  In lattice field theory,  in which Euclidean space-time is discretized in a finite volume, 
with the use of stochastic perturbation theory \cite{Bauer:2011ws,Buividovich:2015qba,Brambilla:2015jiy},  one can in principle extract the large-order asymptotic growth of perturbation theory, and try to find connections with the IR renormalon singularities and resurgence structure.  

 Therefore, under the assumption that resurgence is a phenomenon present in all Quantum Mechanical systems\footnote{There is currently ample evidence indicating that this is true, see the examples  \cite{Dunne:2014bca, Dunne:2013ada} and \cite{CanYuya}. However, the working of the resurgence for quantum systems whose associated classical mechanics is a higher genus Riemann surface is  currently an open problem.}
 one is forced either to concede that the resurgence is operative in both quantum mechanics and quantum field theories, or that the IR and UV regularization of quantum field theories completely change the nature of the theory in question. Since there is much evidence the latter cannot be correct,  (namely,  the very existence and success of lattice field theory is a numerical evidence that  the UV and IR regularization does not alter nature of a QFT dramatically),  we strongly suspect that the resurgence structure is bound to be present in  quantum field theories as well.
 
 Since in our formalism, time is already continuous, and recursion relations happen to have a  geometric (and paralellizable) structure,  in the long run,  we hope that the formalism presented below can be used to determine the connection between perturbation theory and non-perturbative physics in QFT. It is worthwhile checking how efficient is this formalism compared to the conventional path-integral perturbation theory via Feynman diagrams. This is, however, beyond the scope of the present work.

The reduced Hamiltonian we consider is of the form
\be
h=-\sum_{i=1}^{N}\frac{1}{2}\partial_i^2+\frac{1}{g^2}v(gx_1,gx_2,\dots,gx_N)\;.
\ee
with the Schr\"odinger equation
\be
h\psi=\epsilon\psi\;.
\ee
Further, we will assume that 
\be
\frac{v(gx_1,gx_2,\dots,gx_N)}{g^2}=\sum_{i}\frac{\omega_i^2x_i^2}{2}+\tilde v(gx_1,gx_2,\dots,gx_N)\;.
\ee
Using an ansatz
\be
\psi=u(\{x_i\})e^{-\frac{1}{2}\sum_{i=1}^N\omega_ix_i^2}\;,
\ee
we have
\be\label{eq:multi_inter}
\sum_{i=1}^N\left[-\partial_i^2u+2x_i\omega_i\partial_i u\right]+\frac{2}{g^2}\tilde v(\{gx_i\})u=2\left(\epsilon-2\frac{1}{g^2}(0)-\frac{\sum{\omega_i}}{2}\right)u
\ee
Upon writing
\be
u(\{x_i\})=\sum_{l}u_l(\{x_i\})g^l
\ee
and expanding $\tilde v$
\begin{align}
&\tilde v(\{gx_i\}=\sum_{\sum_i{n_i}>2}v_{n_1,n_2,\dots n_N}g^{n_1+n_2\dots+n_N-2}x_1^{n_1}x_2^{n_2}\dots x_N^{n_N}\;,\\
&v_{n_1,\dots n_N}=\frac{1}{n_1!n_2!\dots n_N!}\frac{\partial^{n_1}}{\partial y_1^{n_1}}\dots\frac{\partial^{n_N}}{\partial y_N^{n_N}}\tilde v(\{y_i\})\Bigg|_{\{y_i=0\}}
\end{align}
and
\be
\epsilon=\sum_{n}g^n\epsilon_n\;.
\ee
we obtain from \eqref{eq:multi_inter}
\be\label{eq:multi_inter1}
\sum_{i=1}^N\left[-\partial_i^2u_l+2x_i\omega_i\partial_i u_l\right]+2\sum_{\sum_in_i>2}v_{n_1,n_2,\dots,n_N}u_{l-(n_1+n_2\dots+n_N-2)}x_1^{n_1}x_2^{n_2}\dots x_N^{n_N}=2\sum_{n=0}^l\epsilon_n u_{l-n}\;.
\ee
The equation for $l=0$ 
\be
-\partial_i^2u_0+2x_i\omega_i\partial_i u_0=2\epsilon_0 u_0\;,
\ee
with a solution
\begin{align}
&u_0=H_{\nu_1}(x_1)H_{\nu_2}(x_2)\dots H_{\nu_N}(x_N)\;\\
&\epsilon_0=\nu_1+\nu_2+\dots+\nu_N\;.
\end{align}
which is just the multi-dimensional Harmonic oscillator solution.
Finally we take
\be
u_l=\sum_{k_1,\dots, k_N}A_{l}^{k_1,\dots,k_N}x_1^{k_1}x_2^{k_2}\dots x_N^{k_n}\;,
\ee
and, upon equating powers of $x_i$, obtain
\begin{multline}\label{eq:multi-rec}
\sum_{i=1}^N\left[-(k_i+2)(k_i+1)A_{l}^{k_1,\dots,k_i+2,\dots, k_N}+2\omega_i(k_i-\nu_i)A_{l}^{k_1,\dots k_N}\right]\\+2\sum_{\sum_{i}n_i> 2}A_{l-(n_1+n_2+\dots-2)}^{k_1-n_1,k_2-n_2,\dots, k_N-n_N}v_{n_1,n_2,\dots, n_N}=2\sum_{n}\epsilon_nA_{l-n}^{k_1,k_2,\dots,k_N}\;.
\end{multline}

To solve the recursion relation, we will assume that all the leading order harmonic oscillator states are non-degenerate\footnote{The degenerate case is more subtle, and we postpone it for future work.}, so that the only solution of the equation is
\be
\sum_{i}\omega_i(k_i-\nu_i)=0\Rightarrow k_i=\nu_i\;.
\ee
This allows us to set $k_i=\nu_i$ into the equation \eqref{eq:multi-rec} and obtain
\begin{multline}
\epsilon_l A_0^{\nu_1,\nu_2,\dots,\nu_N}=\sum_{i=1}^N\Bigg[-\frac{1}{2}(\nu_i+2)(\nu_i+1)A_{l}^{\nu_1,\dots,\nu_i+2,\dots, \nu_N}\\-\sum_{n=0}^{l-1} \epsilon_nA_{l-n}^{\nu_1,\nu_2,\dots,\nu_N}+\sum_{\sum_{i}n_i> 2}A_{l-(n_1+n_2+\dots-2)}^{\nu_1-n_1,\nu_2-n_2,\dots, \nu_N-n_N}v_{n_1,n_2,\dots, n_N}\Bigg]
\end{multline}
By choosing the coefficient of $x_1^{\nu_1}x_2^{\nu_2}\dots x_N^{\nu_N}$ to be unity, we may set
\be
A_{0}^{\nu_1,\nu_2,\dots, \nu_N}=1\;,\qquad A_{l}^{\nu_1,\nu_2,\dots,\nu_N}=0\;, \text{when $l\ne0$}\;,
\ee
giving
\be\label{eq:multi-eps-rec}
\epsilon_l =\sum_{i=1}^N\Bigg[-\frac{1}{2}(\nu_i+2)(\nu_i+1)A_{l}^{\nu_1,\dots,\nu_i+2,\dots, \nu_N}+\sum_{\sum_{i}n_i> 2}A_{l-(n_1+n_2+\dots-2)}^{\nu_1-n_1,\nu_2-n_2,\dots, \nu_N-n_N}v_{n_1,n_2,\dots, n_N}\Bigg]\;.
\ee
To find an energy correction $\epsilon_l$ we need to have all $\epsilon_{l'<l}$ and $A_{l'<l}^{k_1,\dots, k_N}$, as well as $A_l^{\nu_1,\dots,\nu_i+2,\dots,\nu_N}$. On the other hand if at least one $k_j\ne \nu_j$ for some $j\in\{1,\dots,N\}$, from \eqref{eq:multi-rec} we obtain
\begin{multline}\label{eq:multi-A-rec}
A_{l}^{k_1,\dots k_N}=\frac{1}{\sum_i\omega_i(k_i-\nu_i)}\Bigg[\sum_{i=1}^N\frac{1}{2}(k_i+2)(k_i+1)A_{l}^{k_1,\dots,k_i+2,\dots k_N}\\-\sum_{n_i> 2}A_{l-(n_1+n_2+\dots-2)}^{k_1-n_1,k_2-n_2,\dots, k_N-n_N}v_{n_1,n_2,\dots, n_N}+\sum_{n=1}^l\epsilon_nA_{l-n}^{k_1,k_2,\dots,k_N}\Bigg]\;.
\end{multline}

To show a way to solve these equations, let us first introduce some notation. Let all the indices $k_1,k_2,\dots, k_N$ be grouped into a vector of integers
\be
\vec k=(k_1,k_2,\dots,k_N)\;.
\ee
Further, let the basis vectors be
\begin{align*}
&\vec e_1=(1,0,\dots,,0)\\
&\vec e_2=(0,1,0,\dots,0)\\
&\vdots\\
&\vec e_N=(0,\dots,0,1)\;.
\end{align*}
Now, let the vector $\vec K_l$ be the maximum power vector at the order $l$
\be
\vec K_l=(K_l^1,K_l^2,\dots, K_l^N)\;,\qquad K_l\in \mathbb N^+\;.
\ee
meaning that
\be
A_l^{\vec k}=0\;,\qquad \text{if  $\exists \vec i\quad$such that$\quad\vec e_i\cdot \vec k>K_l^i$}\;.
\ee
To abbreviate notation, we will denote the condition above
\be
\vec k>\vec K_l
\ee
to mean that \emph{at least one component of the vector $\vec k$ is larger than the vector $s$}. Note that the ordering of this inequality is important, and we have not defined the relation $\vec k<\vec s$.
Notice also that, since the leading order coefficients $A_0^{\vec k}$ are of order no higher than $\nu_1,\nu_2,\dots,\nu_N$ in the coordinates $x_1,\dots, x_N$, the $\vec K_0$ is given by
\be
\vec K_0=(\nu_1,\dots,\nu_N)=\vec \nu\;.
\ee
We can write any vector $\vec k$ as 
\be
\vec k=\vec K_l-\vec s\;.
\ee
Further, we classify values of $\vec k$, uniquely determined by the $\vec s$, by the value of the integer
\be
n_{\vec k}=\sum_{i=1}^N s_n\;.
\ee
We now claim that we can successively solve for the coefficients $A_l^{\vec k}$ for the classes
\be
n_{\vec k}=0,1,2,\dots n_{\vec\nu}
\ee
using the equation \eqref{eq:multi-A-rec}. Indeed notice that since for a given $\vec k$ in a fixed class labeled by the integer $\vec n$, the RHS of \eqref{eq:multi-A-rec} contains only coefficients $A_l^{\vec k+2\vec e_i}, i=1,\dots, N$ which are in the class $n_{\vec k}-2$. Since we assume that we know all the coefficients $A_l^{\vec k'}$ such that $n_{\vec k'}<n_{\vec k}$, we can solve for the all the coefficients $A_{l}^{\vec k}$. However, in order to ensure that $\epsilon_l$ does not appear in the RHS of \eqref{eq:multi-A-rec}, we must restrict ourselves for the coefficients of classes $n_{\vec k}\ge n_{\vec \nu}$. 

Notice that if we know all $A_l^{\vec k}$ with $n_k=0,1,2\dots, n_{\vec \nu}$, we in particular know the class $n_{\vec\nu+2\vec e_1}$. This allows us to use \eqref{eq:multi-eps-rec} to solve for $\epsilon_l$. Now that $\epsilon_l$ is known, we can solve for the classes $n_{\vec k}<n_{\vec \nu}$.

\begin{figure}[t] 
   \centering
   \includegraphics[width=.8\textwidth]{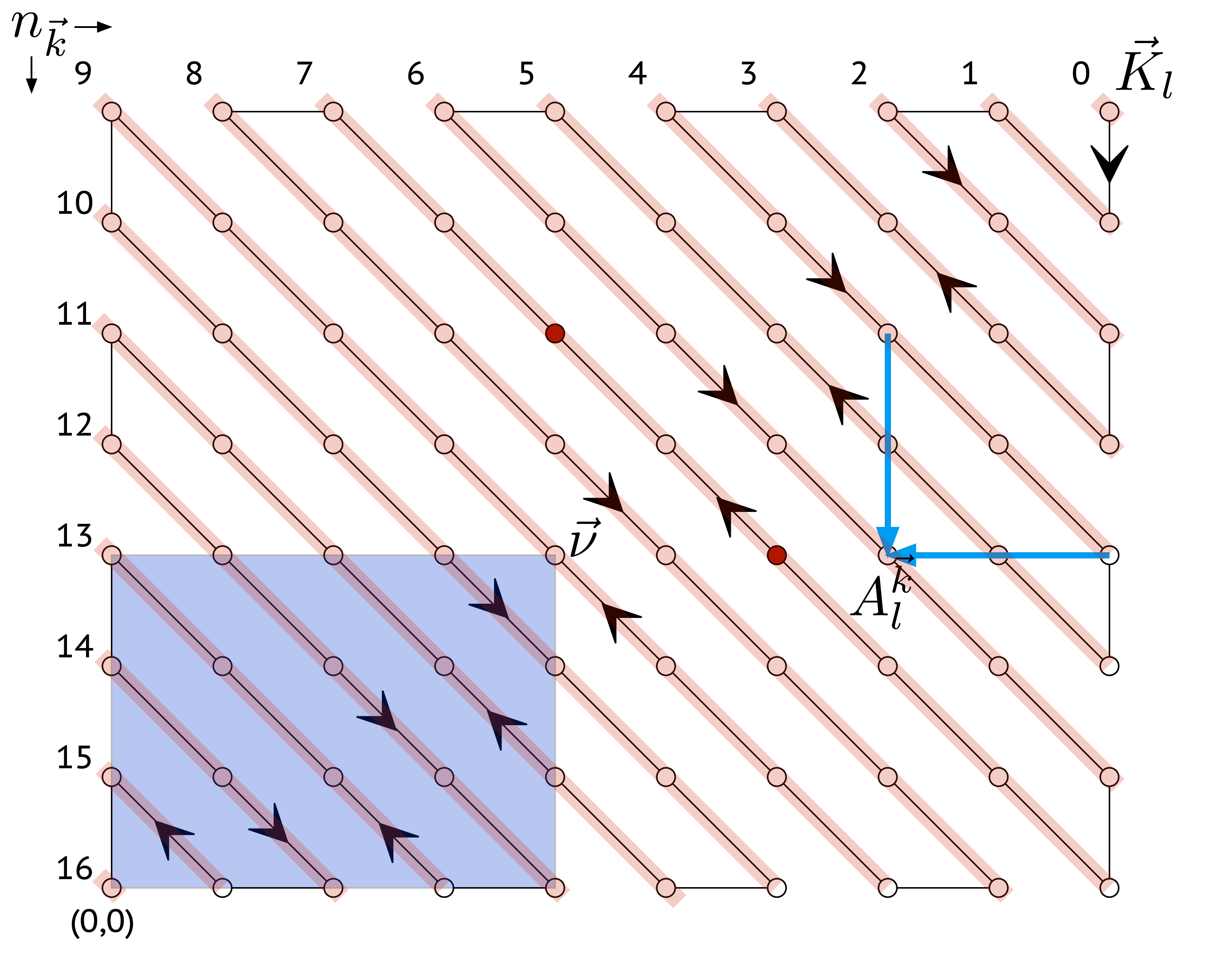} 
   \caption{A recursion diagram showing how all coefficients $A_l^{\vec k}$ can be found using \eqref{eq:multi-A-rec} and \eqref{eq:multi-eps-rec} in the case of the two-variable system.  The white circles denote the vector $\vec k$ for which the coefficients $A_l^{\vec k}$ do not a priori vanish. The red-shading indicates the coefficient $A_l^{\vec k}$ of the same class labeled by the integer $n_{\vec k}$. The black line with an arrow indicates a particular order in which the coefficients $A_l^{\vec k}$ can be solved. The two red circles indicate the value of $\vec k$ for the coefficients $A_l^{\vec k}$ which appear in the equation \eqref{eq:multi-eps-rec} for $\epsilon_l$. The shaded area indicates the values of $\vec k$ for which coefficients $A_l^{\vec k}$ using \eqref{eq:multi-A-rec} can be found only if $\epsilon_l$ is known. The blue arrows indicates that for a coefficient $A_l^{\vec k}$ to be found, the knowledge of  $A_l^{\vec k+2\vec e_1}$ and $A_l^{\vec k+2\vec e_2}$ is required.}
   \label{fig:multi-rec}
\end{figure}

To illustrate this procedure, let us explain how the recursion relations can be solved in the case of the two-variable problem. Then $\vec k=(k_1,k_2)$, and a particular recursion procedure can be presented graphically. We illustrate it in Fig. \ref{fig:multi-rec}. 
Starting from the vector $\vec k=\vec K_l$, we can follow the arrowed path to determine the coefficients $A_l^{\vec k}$. Note \eqref{eq:multi-A-rec} requires the knowledge of the previous coefficients $A_l^{\vec k+\vec e_1}$ and $A_l^{\vec k+\vec e_2}$ (denoted by blue arrows). These coefficients are clearly always located at positions in the diagram for which the $A$-coefficients were already determined\footnote{Note that the successive recursion pattern we show here is not unique. In fact to solve for any coefficient $A_l^{\vec k}$ in the same class $n_{\vec k}$, we need only to know all the coefficients in the class labeled by $n_{\vec k}+2$. This fact allows for the recursion algorithm to be parallelized.}.

\section{The \BenderWu{} \Mathematica{} package}\label{sec:BenderWu}

In this section we describe how the \Mathematica{} \BenderWu{} package is used and what are its abilities. However rather than explaining its features in details, we will mostly focus on working out a particular example. Many more examples are found in an example file accompanying this work, and the reader is encouraged to refer to it. 

\subsection*{Installing and loading the \BenderWu{} package}

To install the \BenderWu{} package, open the accompanying installation \Mathematica{} notebook under the name
\begin{center}
\begin{BVerbatim}
BenderWu_Installation.nb
\end{BVerbatim}
\end{center}
which allows one to easily copy the \BenderWu{} package into your computers repository. Once complete, use the command

\vspace{.5cm}
\noindent\hspace{.5cm}\includegraphics[scale=.28]{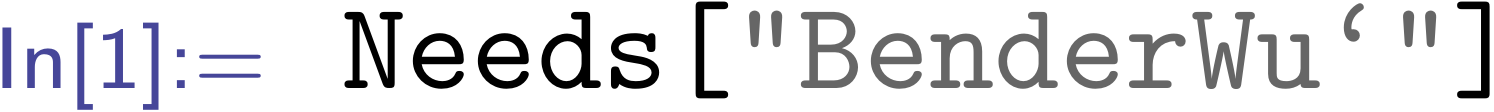}
\vspace{.5cm}

\noindent or

\vspace{.5cm}
\noindent\hspace{.5cm}\includegraphics[scale=.3]{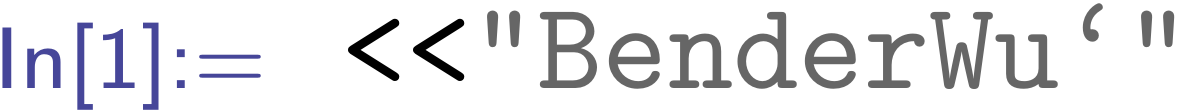}
\vspace{.5cm}

\noindent to load the package. This loads the package content, which comprises the functions: \BenderWu{} function, \BWProcess{} function and \texttt{BWLevelPolynomials}. We briefly describe their application.

\section*{The \BenderWu{} function}

The integral  part of a \BenderWu{} package is the function of the same name. The \BenderWu{} takes in four essential arguments and a number of options. The function is called with the following basic syntax
\begin{verbatim}
BenderWu[V[x],x,(Level Number), (Order),Options]
\end{verbatim}
The arguments above are: 
\begin{itemize}
\item The potential which has a local minimum at $x=0$ (e.g. \verb@x^2+3x^7@)
\item The argument of the potential (e.g. \verb@x@)
\item The level number for which one wishes to compute (e.g. \verb@3@)
\item The order of $g^2$ to which one wishes to compute the energy and the wave-function (e.g. \verb@30@).
\item After fourth argument is called, a number of options can be called. These are discussed in Sec. \ref{sec:options}
\end{itemize}
Notice that the order of $g^2$, and not $g$ which is specified in our normalization\footnote{This labeling is done to be consistent with most of the literature where the order refers to the order of $\hbar\propto g^2$. See the definition of $g$ in the introduction.}.

The first argument, ' can be input in two basic forms: either by just writing the potential, in which case the entry specifies the classical  potential only discussed in Sec. \ref{sec:BWclassical}, or it can be entered as
\begin{verbatim}
BenderWu[{V0[x],V2[x]},x,(Level Number), (Order),Options]
\end{verbatim}
where \verb;V0[x]; and \verb;V2[x]; are the classical and the quantum effective potentials $v_0(x)$ and $v_2(x)$ discussed in Sec. \ref{sec:QEA}.

It is important to remember that the the code assumes that the potential $v_0(x)$  is at a local minimum when the argument is equal to zero. This means that the perturbative analysis will be carried out assuming that $v_0'(x)=0$.  Also it must hold that $v_0''(0)\ne0$. To do an expansion around a minimum at \verb;x=x0;, one can use a syntax 
\begin{verbatim}
BenderWu[{V0[x0+x],V2[x0+x]},x,(Level Number), (Order),Options]
\end{verbatim}

\section*{The \BWProcess{} function}

The \BWProcess{} function serves as a processing function for the output of the \BenderWu{} function. It helps to format the output and present it in a way that is most useful to the user.  It allows for a number of options which control how the output is formatted. The basic syntax is
\begin{verbatim}
BWProcess[(output of BenderWu), Options]
\end{verbatim}

\section*{\texttt{BWLevelPolynomial} function}

The \BenderWu{} function allows only for the insertion of the integer values for the level number, and cannot directly compute the analytical level-number dependence. The \texttt{BWLevelPolynomial} function, however, allows for the reconstruction of the level-number dependence by computing the perturbative data for multiple values of the level-number. The details of how to use this function are found in the Section \ref{sec:BWLevelPolynomial}.

To illustrate how the function \BenderWu{} and \BWProcess{} function we will go through an example of the potential $v(gx)/g^2=\frac{x^2}{2}+g^2 x^4$ explicitly and demonstrate how some of the basic functions and options work. The discussion of the use of the usage of the function \texttt{BWLevelPolynomial} is postponed to the section \ref{sec:BWLevelPolynomial}. In addition more examples have been worked out in the \Mathematica{} notebook \verb@Examples.nb@ accompanying this work. The reader is strongly encouraged to refer to this supplemental material.

\subsection{An example}\label{sec:BW_example}

Let us consider an example where the \BenderWu{} function is used to compute the first 100 corrections of of the perturbation for the third excited state for the potential $v(xg)/g^2=\frac{x^2}{2}+g^2x^4$, where $g$ is treated as an expansion parameter:

\vspace{.5cm}
\noindent\hspace{.5cm}\includegraphics[scale=.3]{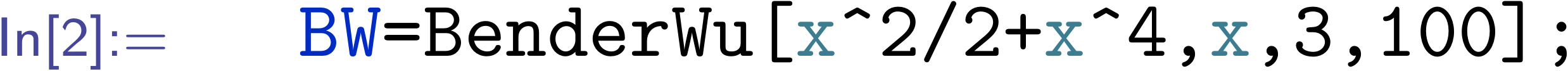}
\vspace{.5cm}

Upon execution of the above command, a  progress monitor will appear indicating some basic information about the progress of the computation. An example of the progress monitor is given below:

\vspace{.5cm}
\hspace{1cm}\includegraphics[scale=.3]{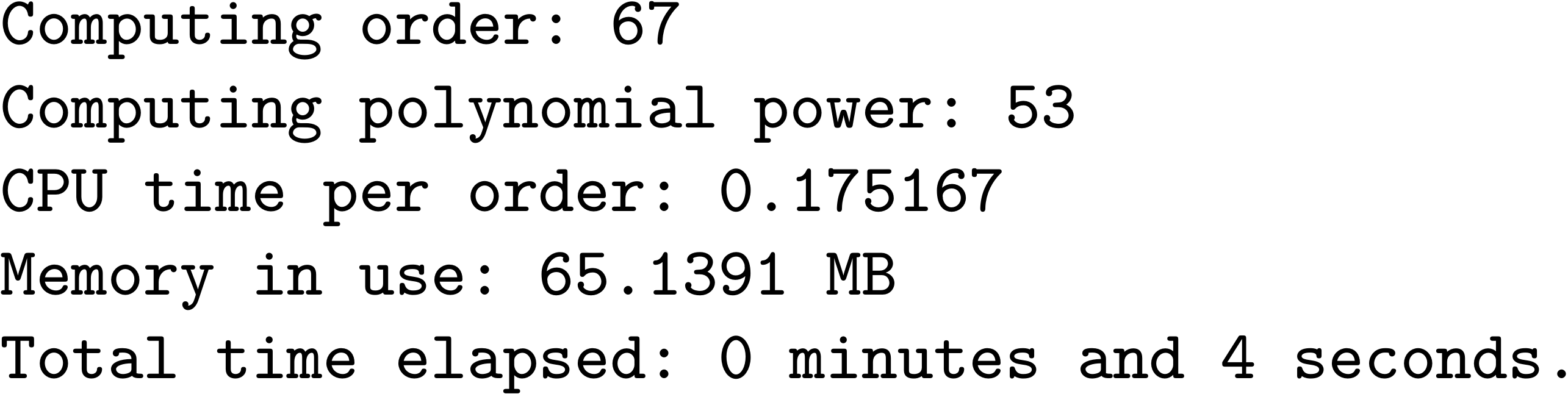}
\vspace{.5cm}
The progress monitor gives the information about:
\begin{itemize}
\item The order of $g^2$ it is currently computing, the CPU time it needs per order,
\item The  power of $x$ (which we labeled as ``$k$'' in Sec. \ref{sec:BW})  it is currently computing ,
\item The CPU time it takes for each order,
\item The total memory in use,
\item The total time elapsed since the beginning of the computation.
\end{itemize}

In certain situations it may be desirable to turn off the Progress Monitor. To do that use the option \verb@Monitor->False@, e.g. 

\vspace{.5cm}
\noindent\hspace{.5cm}\includegraphics[scale=.3]{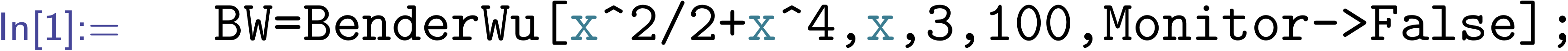}
\vspace{.5cm}

Once the computation is done, the symbol \verb@BW@ is assigned an array with three elements. The elements of the array are, in order: energy corrections $\epsilon_l$, wave-function coefficients $A_l^k$ and a list of 

The first element of the variable \verb@BW@ gives a series of coefficients.

\vspace{.5cm}
\noindent\hspace{.5cm}\includegraphics[scale=.3]{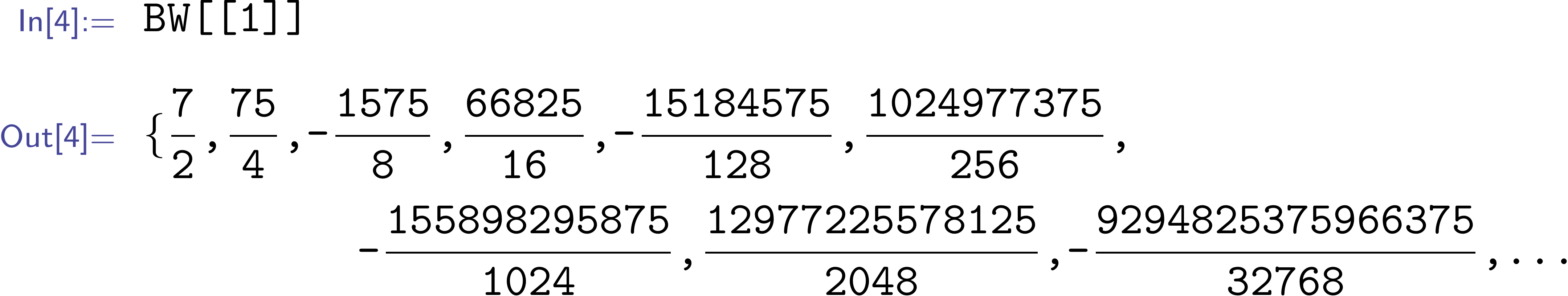}
\vspace{.5cm}

\noindent where we indicated with dots that we have trimmed the output as it is too long at this order.
The second element, \verb@BW[[2]]@ outputs a matrix $A_{l}^k$ where rows are associated with index $l$ (the order of $g$) and $k$ is associated with the index $k$ (the power of $x$). Since this output is large and not very illuminating, we will not show it here. We will see in a moment how one can use post-processing function \verb@BWProcess@ in order to generate a more convenient output.

The function \verb@BWProcess@ takes in the default output of the function \BenderWu{}. It allows the raw data generated by the call to of \BenderWu{} function to be post-processed into an output which is most suitable for the user. If called without any options it returns the energy eigenvalues.

\vspace{.5cm}
\noindent\hspace{.5cm}\includegraphics[scale=.3]{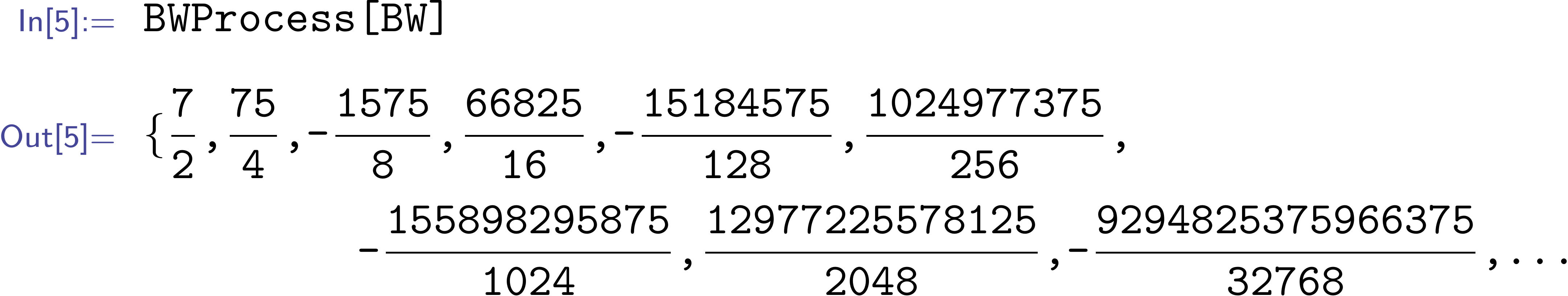}
\vspace{.5cm}

\noindent However, to change the output into a series in $g$, we can simply use the option \verb@OutputStyle->"Series"@

\vspace{.5cm}
\noindent\hspace{.5cm}\includegraphics[scale=.3]{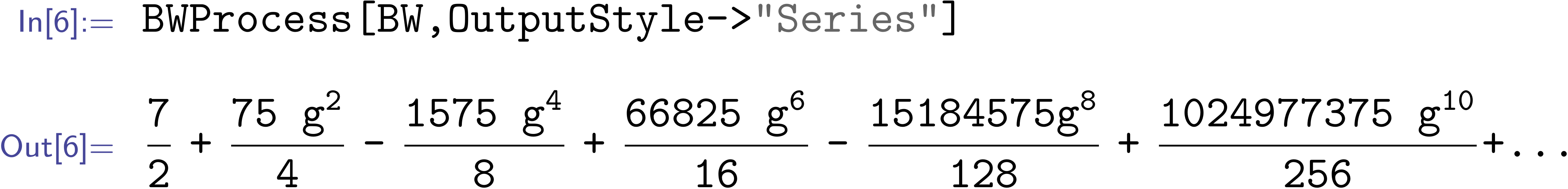}
\vspace{.5cm}

\noindent where again we indicated with dots that we have trimmed the output because it is too large.
Indeed, often the full output is too long to be illuminating, and one is most interested in the first few orders. 
To generate an output which only contains data up to a specific order, one can use the option  \verb@Order->n@, where \verb@n@ specifies the order (of $g^2$) at which to trim the output, e.g.  

\vspace{.5cm}
\noindent\hspace{.5cm}\includegraphics[scale=.3]{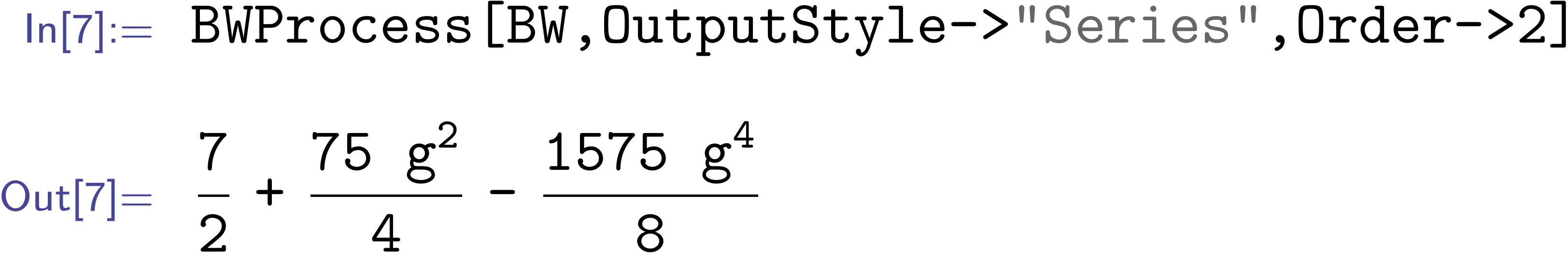}
\vspace{.5cm}

We can also output the wave-function series $u(x)=\sum_{l,k}A_{l}^kg^l x^k$ by utilizing the option \newline\texttt{Output->"WaveFunction"}. This gives

\vspace{.5cm}
\noindent\hspace{.5cm}\includegraphics[scale=.3]{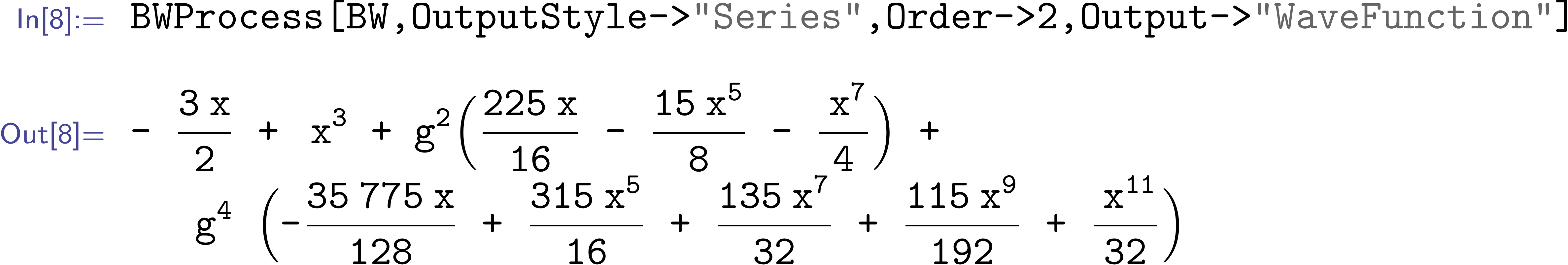}
\vspace{.5cm}

\noindent Note that the options \verb@Output@ and \verb@OutputStyle@ of the \verb@BWProcess@ function are also the options of \BenderWu{}. This is also true for many other options, discussed in the next section. Therefore, the last output we generated could have equally been generated by calling the \BenderWu{} function like so

\vspace{.5cm}
\noindent\hspace{.5cm}\includegraphics[scale=.3]{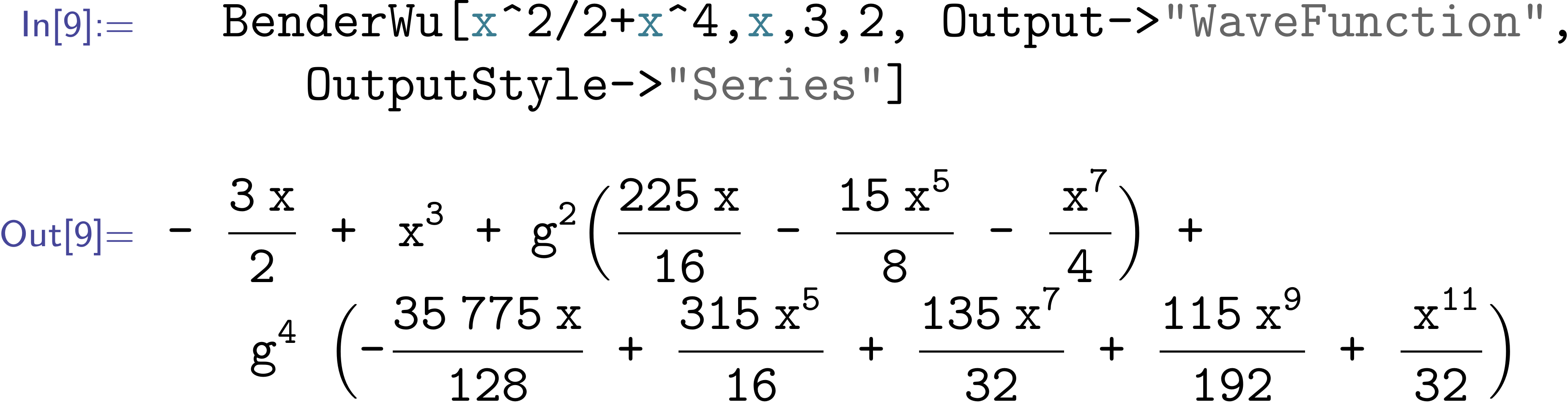}
\vspace{.5cm}

\noindent Note finally that the potential can also contain an arbitrary non-numerical parameter, i.e. 

\vspace{.5cm}
\noindent\hspace{.5cm}\includegraphics[scale=.3]{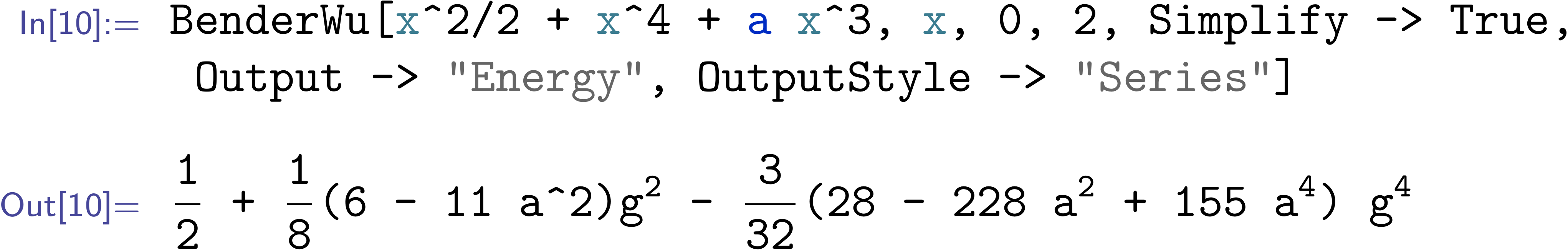}
\vspace{.5cm}

\noindent Notice that we also used the option \texttt{Simplify->True}. This option simplifies the analytical expressions at each step, often resulting in faster evaluation an a simplified result when non-numerical symbols are used, as well as when irrational factors appear in the calculation.
\subsection{The options}\label{sec:options}

The \BenderWu{} function can be called without only the four essential arguments. The output is then given as an array with three entries. The first entry is an array of perturbative corrections to the energies $\epsilon_k$. The second entry is a matrix $A_l^k$ of the wave-function coefficients. The third entry passes arguments $x,\nu,l_{max}$ as well as a variable \verb@even@ which determines whether the function is even or not. The purpose of the third entry in the output array is to allow the post-processing by the function \BWProcess{} which uses these values.

The basic evaluation options which can be specified in the \BenderWu{} function are given in the Table \ref{tab:eval_options}, as well as a brief description.

The \BenderWu{} also allows for a number of options which determine the way the data is output. These options are options of the function \BWProcess{} which takes in a single argument

The options are called with the standard \Mathematica{} syntax: 
\begin{center}
\begin{BVerbatim}
(Option)->(Option Value)
\end{BVerbatim}
\end{center}

\noindent The table of options of as well as their description is given in tables \ref{tab:eval_options} and \ref{tab:proc_options}

\begin{table}\scriptsize\label{tab:eval_options}
\setlength\tabcolsep{5pt}
\setlength{\extrarowheight}{5pt}
\begin{tabular}{l | l | p{8cm}} 
\textbf{Option} & \textbf{Values} & \textbf{Description}\\
\hline\hline 
\verb@Evaluation@		&\verb@"Analytical" (default)@ &Use symbolic calculation\\
					&\verb@"Numerical"@ &Evaluate numerically (good if irrational constants appear in the potential or harmonic oscillator frequency)\\\hline
\verb@WorkingPrecision@&\verb@MachinePrecision@ (default)& Use hardware machine precision for numerical evaluation (recommended)\\
					&non-negative integer& The number of digits used for internal computations\\[5pt]\hline
\verb@MaxTime@&Infinity (default)& Perform a computation to the specified order, regardless of the length of time it takes.\\
					&number $>0$& Specifies the maximal number of seconds used for a computation of a single order\\\hline
\verb@TotalMaxTime@&Infinity (default)& Perform a computation to the specified order, regardless of the length of total time it takes.\\
					&number $>0$& Specifies the maximal number of minutes used for an entire computation\\\hline
\verb@Simplify@		&\texttt{False} (default) or \texttt{True} & If \texttt{True} then the function \texttt{Simplify} is used in each step of the evaluation. \\\hline
\verb@Monitor@		 & \verb@True@ (default) & Turns on the monitor during the computation\\
							 & \verb@False@ (default) & Turns off the monitor during the computation\\
\end{tabular}
\caption{The evaluation options of the \BenderWu{} function. }
\end{table}

\begin{table}\scriptsize\label{tab:proc_options}
\setlength\tabcolsep{5pt}
\setlength{\extrarowheight}{5pt}
\begin{tabular}{l | l | p{6cm}}
\textbf{Option} & \textbf{Values} & \textbf{Description}\\
\hline\hline 
\verb;Output; 			& \verb;"Energy" (default for {BWProcess}); & Outputs the energy data $\epsilon_l$\\
					&\verb;"WaveFunction";& Outputs the wave-function coefficients $A_l^k$\\\hline
\verb@OutputStyle@ 	&\verb@"Array"@ (default) &Outputs the data as an array\\
				 	&\verb@"Series"@ &Outputs the data as a series using "g" as the series coefficient (see also option \verb@Coupling@ and \verb@SeriesOutput@)\\
					&\verb@"MatrixForm"@ &Outputs the data as a matrix \\
					&\verb@"Ratio"@ &Affects only the "Energy" output. Generates a list of ratios of the series coefficients $a_{n+1}/a_{n}$.\\
					&\verb@"RatioPlot"@ &Affects only the "Energy" output. Generates a \texttt{ListPlot} of the ratios of the coefficients $a_{n+1}/a_n$.\\\hline
\verb@SeriesOutput@	&\verb@"Coupling" (default)@	&If series output is used the terms are grouped according to the coupling\\
					&\verb@"Argument"@			&If series output is used, the terms are grouped according to the argument $x$\\\hline
\verb@Coupling@& g (default) &Specifies a symbol used for the coupling in the series output\\[5pt]\hline
\verb@Order@	(\BWProcess{} only)	&0 (default) or $n\in$Integer & If called with integer other than zero, then it truncates the data to the order $n$.\\
							& $\{n_{min},n_{max}\}$ & It produces an output from order $n_{min}$ to order $n_{max}$. The order $n_{min}$ can take values $-1,0,1,\dots$. If $n_min=-1$ the classical shift is included in the output for the energy.

\end{tabular}
\caption{The output formatting options of the \BenderWu{}  and the \BWProcess{} function. All of the options, save for the option \texttt{Order} are shared between the two functions. Notice that if no options are called in the \BenderWu{} function, or only evaluation options of table \ref{tab:proc_options} are called, the default output is in terms of raw data which serves as an input to the \BWProcess{} function.}
\end{table}

\subsection{The polynomials in level number and the \texttt{BWLevelPolynomial} function}\label{sec:BWLevelPolynomial}

In this section we describe another function of the package \BenderWu{} which allows one to compute the functional form of the energy correction $\epsilon_{2n}$ (recall that we call $n$ \emph{the order}) on the level number $\nu$ , assuming that it is in the polynomial form. In other words we will assume that
\be
\epsilon_{2n}={a_0}^n+{a_{1}}^n \nu+{a_{2}}^n \nu^2+\dots+{a_{p_{p}}}^n \nu^{p(n)}
\ee
where $p$ is the highest power of $\nu$ that appears. Notice that $p$ depends on $n$ in general.

To determine the polynomial form we must
\begin{itemize}
\item Assume that at the maximal order $n_{\text{max}}$ we are interested in, the highest power of the polynomial in $\nu$ is $p_{\text{max}}=p(n_{\text{max}})$.
\item Use the \BenderWu{} function to compute the energy corrections up to the order $n_{\text{max}}$ and for values of $\nu=0,1,2,\dots \nu_{\text{max}}$, where $\nu_{\text{max}}\ge p_{n_{\text{max}}}$.
\item Use the data to fix the coefficients  ${a_{i}}^n$ ($n=0,1,\dots,n_{\text{max}},i=0,1,\dots, p(n)$) of a polynomial in $\nu$.
\end{itemize}

The function \texttt{BWLevelPolynomial} does precisely that. It takes a two-dimensional array, with first-level entries of the energy series coefficients $\epsilon_{2n}$ up to some order $n_{\text{max}}$, for various level numbers $\nu$, and returns an array of the polynomials in $\nu$ corresponding to each order of $n=0,\dots n_{\text{max}}$. 

In other words, by taking $\nu_{\text{max}}=4$ and $p_{\text{max}}=3$  we can simply generate a table

\vspace{.5cm}
\noindent\hspace{.5cm}\includegraphics[scale=.3]{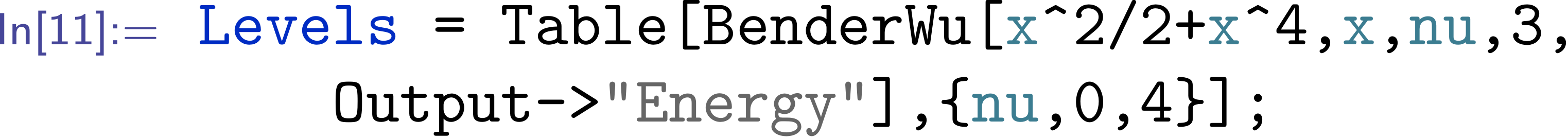}
\vspace{.5cm}

\noindent which produces a two dimensional array, or a matrix $e_{2n}(\nu)$, where $n$ index is the column index and $\nu$ is the row index. This matrix can be forwarded to the function \texttt{BWLevelPolynomial} to produce an output

\vspace{.5cm}
\noindent\hspace{.5cm}\includegraphics[scale=.3]{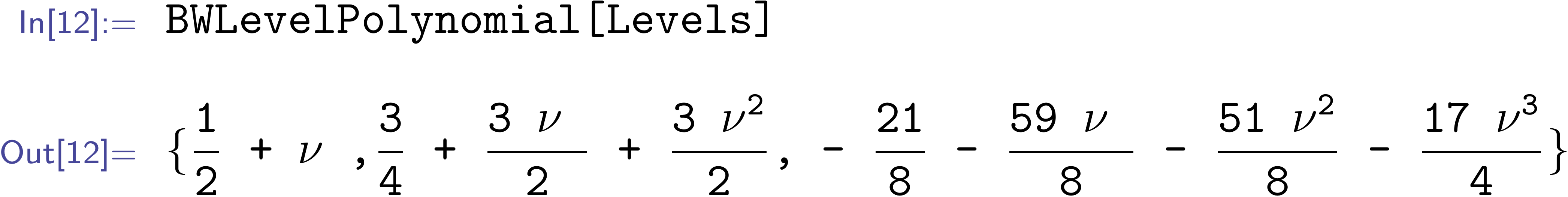}
\vspace{.5cm}

\noindent returning the polynomial form of the three orders which we computed. Notice that the leading order is just the usual Harmonic Oscillator term, while the rest correspond to the Bohr-Sommerfeld quantization terms. The levels function can also take an optional argument which is used as a symbol for $\nu$. 

\vspace{.5cm}
\noindent\hspace{.5cm}\includegraphics[scale=.3]{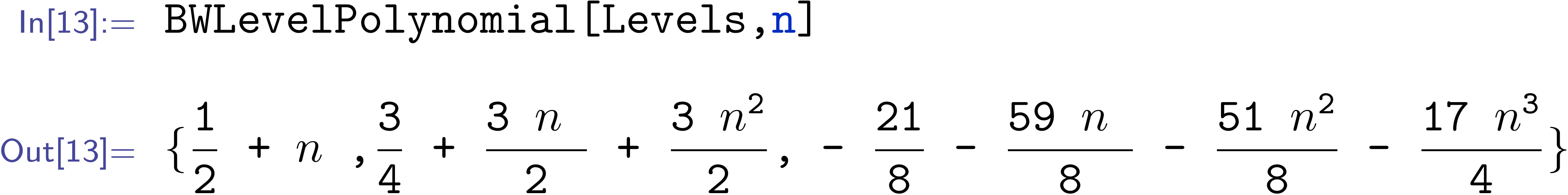}
\vspace{.5cm}

\noindent Note that, although technically we can have $\nu_{\text{max}}=p_{\text{max}}$, since $p_{\max}$ is not a priori known for the function \texttt{BWLevelPolynomial} to be able to check that the returned polynomials are correct, it needs extra data. Therefore it is necessary that $\nu_{\text{max}}> p_{\text{max}}$. Running the same steps, except taking $\nu_{\text{max}}=3$ results in a warning and a truncated result

\vspace{.5cm}
\noindent\hspace{.5cm}\includegraphics[scale=.3]{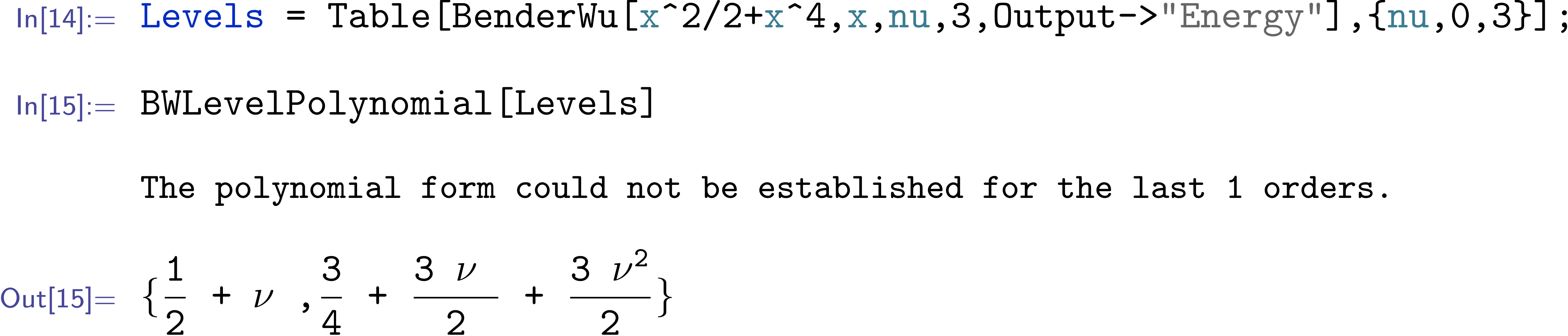}
\vspace{.5cm}

\noindent returning only the results which were computed reliably.

\section{Conclusion}\label{sec:conclusions}

We have presented the generalization of the Bender-Wu recursion relations treating all locally harmonic potentials uniformly. In addition we considered various effective action forms, and derived generalizations for them. 

Additionally we have developed a \Mathematica package incorporating these relations for an easy computation of the perturbative expansions in 1D quantum mechanics. Although the method of Bender-Wu is well known in the literature, the application seems to have been focused on a case-by-case basis. It is our hope that the unified treatment, and simplicity of the \Mathematica package introduced here will stimulate a more fervent development of the resurgence program, as well as bring the program closer to a wider audience. We further hope to adapt the package into a Wolfram demonstration project, accessible to students and researchers who  currently do not have access to a \Mathematica license.

 We also  hope to develop the package to include new, as well as incorporate already known methods of processing and analyzing the resurgence data. Some immediate obvious generalizations are incorporations of arbitrary effective actions, ability to easily change the initial harmonic oscillator ansatz incorporating an analytically resummed subclass (see \cite{Bender:1996ti}), and generalizations to multi-variable Schr\"odinger equation. The last generalization is of potentially vital importance in the attempt to connect resurgence ideas to quantum field theories.

\section*{Acknowledgments}

We would like to give special thanks to Aleksey Cherman for thoroughly testing the code and giving useful comments on an early draft of this work. In addition we are very thankful to Gerald Dunne, Can Koz\c caz and Yuya Tanizaki for discussions, suggestions and code-testing.  This research was supported by the DOE under Grants No.~DE-FG02-03ER41260 (TS) and DE-SC0013036 (M\"U).

\newpage
\begin{appendix}

\section{The highest power for $v_{1}=v_2=\dots=v_{L-1}=0$ and $v_{L}\ne0$.}\label{app:highest_power}

Here we prove the formula 
\be\label{eq:K_l_formula}
K_{l}=\begin{cases}\nu \delta_{l0} &\text{for $0\le l<L$}\\
\nu+(L+2)\left\lfloor \frac{l}{L}\right\rfloor+(l\bmod L)& \text{for $l\ge L$}
\end{cases}
\ee
We will give two proofs. The first one is more intuitive and uses the standard textbook methods of Reyleigh-Schr\"odinger, while the second one is using the Bender-Wu recurrence relations. 

\subsection{The Rayleigh-Schr\"odinger proof}
Consider the Schr\"odinger equation
\be\label{eq:schro_ket}
(h_0+v_{I})\ket{\psi}=\epsilon\ket\psi
\ee
where we separated the Hamiltonian of \eqref{eq:gen_schro} into the harmonic oscillator part $h_0$ and the ``interaction'' terms 
\begin{align}
&h_0=-\frac{1}{2}\partial_x^2+\frac{1}{2}\omega^2x^2\\\label{eq:v_I}
&v_{I}=\frac{\tilde v(gx)}{g^2}=v_1gx^3+v_2g^2x^4+\dots\;.
\end{align}
Now write $\epsilon=\epsilon_0(\nu)+\tilde\epsilon$. The Schr\"odinger equation \eqref{eq:schro_ket} can be formally rewritten as
\be
\ket\psi=\ket \nu+\frac{1}{h_0-\epsilon_0(\nu)}P(\tilde\epsilon-v_{I})\ket\psi
\ee
where $\ket \nu$ is the harmonic oscillator state at level $\nu$, and $P$ is the projector onto the subspace orthogonal to $\ket \nu$. Notice that when $v_I\rightarrow0$ then $\tilde\epsilon\rightarrow 0$ and $\ket\psi\rightarrow \ket \nu$, as it should. 

Now we can iterate the above equation by plugging in the LHS to the RHS. We then get
\begin{equation}\label{eq:iter_sol}
\ket\psi=\ket \nu+\frac{1}{h_0-\epsilon_0(\nu)}P(-v_{I})\ket \nu+\frac{1}{h_0-\epsilon_0(\nu)}P(\tilde\epsilon-v_{I})\frac{1}{h_0-\epsilon_0(\nu)}P(-v_{I})\ket \nu+\dots
\end{equation}
We can now insert complete set of Harmonic oscillator states, and obtain that the  $k$-th power contribution of of $v_I$ is of the form
\be
\psi^{(k)}(x)\propto \sum_{\{n_i\}}H_{n_1}(x)e^{-\frac{\omega x^2}{2}}\frac{v_{n_1n_2}}{\epsilon_{n_1n_2}}\frac{v_{n_2n_3}}{\epsilon_{n_2n_3}}\dots \frac{v_{n_k\nu}}{\epsilon_{n_kn_\nu}}
\ee
where we specified to the coordinate basis, and where
\be
v_{n_{i}n_{i+1}}=\bra{n_i}v_{int}\ket{n_{i+1}}\;,\qquad \epsilon_{n_in_{i+1}}=\epsilon_0({n_i})-\epsilon_0({n_{i+1}})\;.
\ee
The $H_{n}(x)$ are Hermite polynomials of order $n$. If $v_1\ne 0$ in \eqref{eq:v_I}, then by replacing $v_I$ with its leading term\footnote{It can be shown by staring at the equations that this term maximizes the power of $x$ at order $g^{k}.$} $v_1g x^3$ would allow $n_1$ at order $g^{k}$ to be at most $\max[n_1]=\nu+3k$, because each power of $v_I$ would contribute an $x^3$ term, and change the harmonic oscillator level number by $3$. This confirms \eqref{eq:K_l_formula} for $L=1$. 

Now if $v_i=0,i<L$ and $v_L\ne 0$ then the lowest power of $g$ in the potential $v_I$ is $g^L v_L x^{L+2}$, so that there is no contribution to the wave-function until order $g^{L}$.  Therefore all $K_l=0$ for $l=1,2,\dots L-1$. Now for order $g^{L}$, it is easy to see $K_L=K_0+L+2=\nu+L+2$. At order $g^{L+1}$ we can take the term in $v_I$  proportional to $v_{L+1}g^{L+1}x^{L+3}$, and get $K_{L+1}=\nu+L+3$. Continuing one can show that formula \eqref{eq:K_l_formula} is indeed valid.

\subsection{The proof by the recurrence equations}

The proof is by induction. 
First notice that if we take $l<L$ then the formula \eqref{eq:Arec-1} becomes
\be\label{eq:Arec_l_lt_L}
A_{l}^k=\frac{1}{2\omega(k-\nu)}\left[(k+2)(k+1)A_l^{k+2}+2\epsilon_1 A_{l-1}^k+2\epsilon_2 A_{l-2}^k+\dots 2\epsilon_l A_0^k\right]
\ee
while \eqref{eq:eps-l} becomes 
\be\label{eq:eps_l_lt_L}
\epsilon_l=-\frac{1}{2}(\nu+2)(\nu+1)A_{l}^{\nu+2}\;.
\ee
It is easy to convince oneself that these two equations imply $\epsilon_l=0$ and $A_{l}^k=0$ for $l<L$. This just means that there is no corrections to the Harmonic oscillator states and energies up to order $g^{L}$, which is the first order at which the potential gives a non-trivial contribution. This translates into 
\be
K_l=0,\qquad \text{for }0<l<L\;.
\ee
So the first contribution comes at order $l=L$. Plugging that into \eqref{eq:Arec-1}, we get
\be
A_{L}^k=\frac{1}{2\omega(k-\nu)}\left[(k+2)(k+1)A_L^{k+2}-2v_L A_{0}^{k-L-2} \right]\;.
\ee
Since $A_{0}^{k-L-2}=0$ for $k>K_0+L+2=\nu+L+2$, then it must follow that
\be
K_L=\nu+L+2\;.
\ee
Now consider $l=L+1$. Then \eqref{eq:eps-l} becomes
\be
A_{L+1}^k=\frac{1}{2\omega(k-\nu)}\left[(k+2)(k+1)A_{L+1}^{k+2}-2v_LA_1^{k-L-2}-2v_{L+1}A_{0}^{k-L-3}\right]\;.
\ee
The two last terms vanish if both $k>K_1+L+2=2(L+2)$ and $k>K_0+L+3=\nu+(L+2)+1$, so 
\be\label{eq:K_Lp1}
K_{L+1}=\nu+L+3\;.
\ee
Now for $L\le l<2L$, we have
\begin{multline}
A_{l}^k=\frac{1}{2\omega(k-\nu)}\Big[(k+2)(k+1)A_{L+1}^{k+2}\\
+2\epsilon_{L}A_{l-L}^k+2\epsilon_{L+1}A_{l-L-1}^k+\dots+2\epsilon_{l}A_0^k
\\-2v_LA_{l-L}^{k-L-2}-2v_{L+1}A_{l-L-1}^{k-L-3}\dots-2v_{l-1}A_{1}^{k-l-1}-2v_{l}A_{0}^{k-l-2}\Big]\;,
\end{multline}
Now notice that all the terms, except the first on the RHS vanish if $k>K_0+l+2=\nu+l+2$, so that
\be
K_l=\nu+(l+2)\;, L\le l<2L\;,
\ee 
which agrees with \eqref{eq:K_Lp1} when $l=L+1$. 

Now that we have proven that the formula \eqref{eq:K_l_formula}  is valid for $l<2L$, let us assume that it is valid for all $l<\tilde l$. 
Now notice that the formula \eqref{eq:K_l_formula} implies
\be\label{eq:GE_form}
K_{\tilde l}\ge K_{\tilde l-1}+1\;.
\ee
Now taking $l=\tilde l$, the last term in the square bracket of \eqref{eq:Arec-1}
\be
v_L A_{\tilde l-L}^{k-L-2}+v_{L+1}A_{\tilde l-L-1}^{k-L-3}+v_{L+2}A_{\tilde l-L-2}^{k-L-4}+\dots v_{\tilde l-1}A_1^{k-\tilde l-1}+v_{\tilde l} A_0^{k-\tilde l-2}
\ee
will dictate the maximum $k$ for which $A_{\tilde l}^k$ does not vanish. In other words, in order for $A_l^k$ to vanish, all terms above must vanish as well. It is clear that in order for this to be fulfilled, we must have that
\begin{multline}
\Big[k>K_{\tilde l-L}+(L+2)\Big]\;\&\; \Big[k>K_{\tilde l-L-1}+(L+3)\Big]\;\&\; \Big[k>K_{\tilde l-L-2}+(L+4)\Big]\;\&\;\\\dots \;\&\; \Big[k>K_{1}+(\tilde l+1)\Big]\;\&\; \Big[k>K_{0}+(\tilde l+2)\Big]
\end{multline}
\end{appendix}
However, notice that because \eqref{eq:GE_form} holds for all $l<\tilde l$, all of the above conditions are fulfilled if the first one is fulfilled. Therefore
\be
K_{\tilde l}=K_{\tilde l-L}+(L+2)=\nu+(L+2)\left(1+\left\lfloor \frac{\tilde l-L}{L}\right\rfloor \right)+(\tilde l\bmod L)=\nu+(L+2)\left\lfloor \frac{\tilde l}{L}\right\rfloor+(\tilde l\bmod L)
\ee
which implies that formula \eqref{eq:K_l_formula} is also true for $l=\tilde l$.  This finally proves \eqref{eq:K_l_formula} for all $l$.

\bibliography{bibliography}
\bibliographystyle{JHEP}

 \end{document}